\documentclass[twocolumn,aps,showpacs,pre,amsmath]{revtex4-1}

\usepackage{amssymb}
\usepackage{graphicx}

\begin{document}

\title{Improved performance of graphene transistors by strain engineering}

\author{V. Hung Nguyen$^{1,2}$\footnote{E-mail: hung@iop.vast.ac.vn}, Huy-Viet Nguyen$^{2}$ and P. Dollfus$^3$}
\address{$^1$L$_-$Sim, SP2M, UMR-E CEA/UJF-Grenoble 1, INAC, 38054 Grenoble, France \\ $^2$Center for Computational Physics, Institute of Physics, Vietnam Academy of Science and Technology, P.O. Box 429 Bo Ho, 10000 Hanoi, Vietnam \\ $^3$Institut d'Electronique Fondamentale, UMR8622, CNRS, Universit$\acute{e}$ Paris Sud, 91405 Orsay, France}

\begin{abstract}
By means of numerical simulation, we study in this work the effects of uniaxial strain on transport properties of strained graphene heterojunctions and explore the possibility to achieve good performance of graphene transistors using these hetero-channels. It is shown that a finite conduction-gap can open in the strain junctions due to the strain-induced deformation of graphene bandstructure. These hetero-channels are then demonstrated to improve significantly the operation of graphene field-effect-transistors (FETs). In particular, ON/OFF current ratio can reach a value of over 10$^5$. In graphene normal FETs, transconductance, though reduced compared to the case of unstrained devices, is still high while good saturation of current can be obtained. This results in high voltage gain and high transition frequency of a few hundreds of GHz for a gate length of 80 nm. In graphene tunneling FETs, subthreshold swing lower than 30 $mV/dec$, strong non-linear effects such as gate controllable negative differential conductance, and current rectification are observed.
\end{abstract}

\maketitle

\section{Introduction}

Graphene, due to its outstanding electronic properties \cite{ahcn09}, has been expected to be an excellent candidate for advanced applications in future electronics \cite{schw10,ywu013}. In particular, graphene field-effect transistors (FETs) take advantage of high carrier mobility \cite{bolo08,zome11} and high critical current density \cite{novo04}, which makes them suitable for operating in high-frequency ranges \cite{schw10}. Intrinsic cut-off frequency of a few hundreds of GHz \cite{ywu012,chen12} and a record maximum oscillation frequency of 70 GHz \cite{zguo13} in graphene devices have been reported recently. However, for practical applications, graphene transistors still have serious drawbacks associated with the lack of energy bandgap and a poor saturation of current \cite{schw10}. The former makes it difficult to turn off the current, leading to a small ON/OFF current ratio typically lower than ten \cite{meri08}. The latter results in a small voltage gain and hence in the power loss when the devices are integrated in a circuit. So far, many efforts of bandgap engineering in graphene \cite{yhan07,khar11,lher13,jbai10,zhan09} have been made to overcome these limitations. A common technique consists in cutting 2D graphene sheets into 1D narrow nanoribons \cite{yhan07}. To open a bandgap in 2D graphene sheets, some prominent options are Bernal-stacking of graphene on hexagonal boron nitride substrate \cite{khar11}, nitrogen-doped graphene \cite{lher13}, graphene nanomesh lattice \cite{jbai10} and Bernal-stacking bilayer graphene \cite{zhan09}. However, each technique has its own issues. For instance, graphene nanoribbon (GNR) devices require a narrow width with a good control of edge disorder \cite{quer08}. Graphene nanomesh lattices have the same requirement about the good control of the lattice of nanoholes and the disorder \cite{hung13}, while the bandgap opening in bilayer graphene may not be large enough to achieve a sufficiently high ON/OFF ratio in transistors for digital applications \cite{fior09}. Other proposed methods still need experimental verification and realization.

\begin{figure}[!t]
\centering
\includegraphics[width=3.0in]{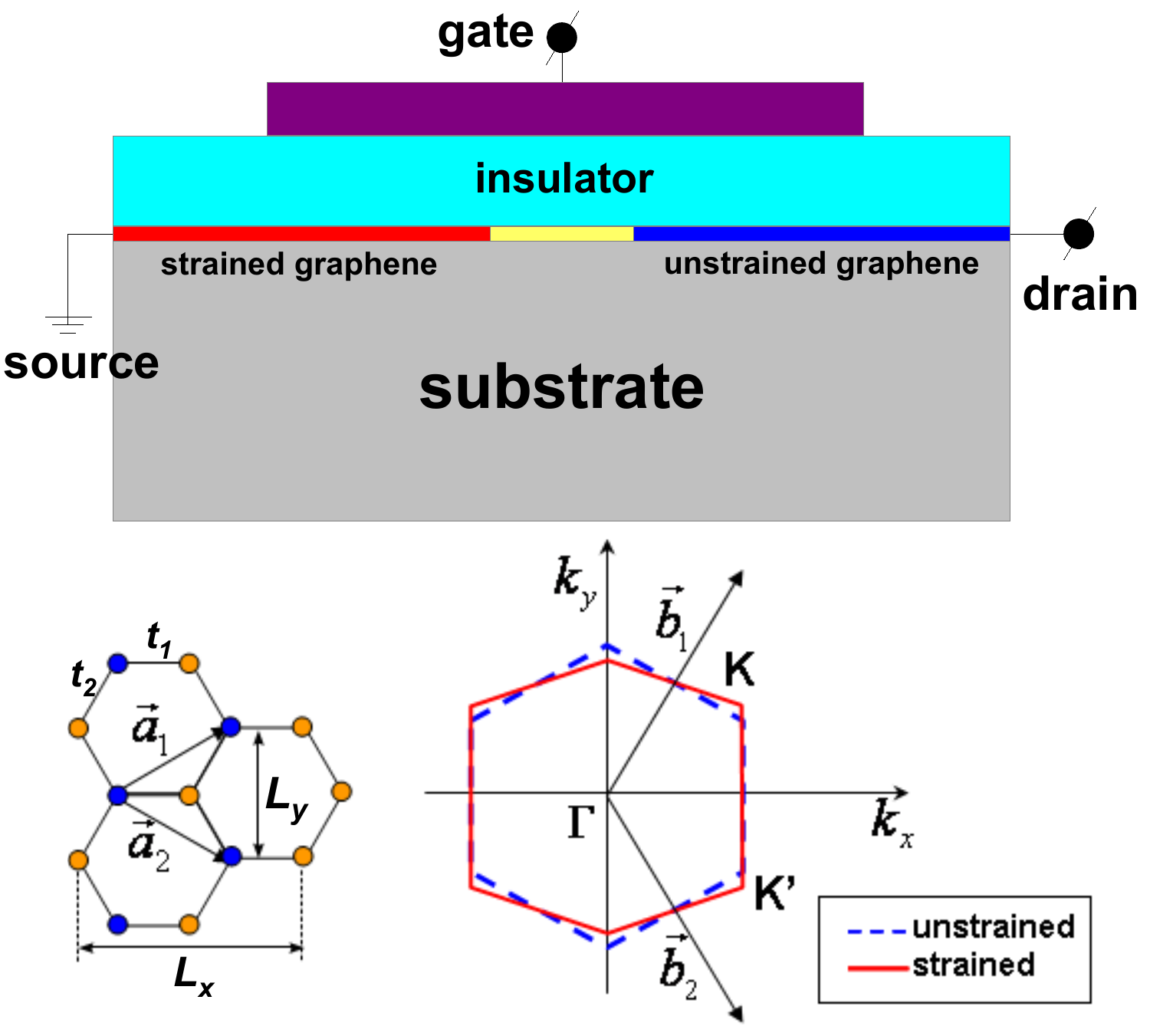}
\caption{Schematic of graphene FETs simulated in this work (top) and graphene lattice together with its first Brillouin zone (bottom). The channel of the simulated device is based on strained graphene heterostructures. The yellow zone in the top figure indicates the transition region between strained and unstrained sections. In the bottom, the left is the honeycomb structure with the lattice vectors \textbf{a}$_1$ and \textbf{a}$_2$. $L_{x,y}$ are the lengths of unit cell (four atoms per cell) along the Ox and Oy directions, respectively. On the right are the Brillouin zones showing the change in Dirac points (\textbf{K} and \textbf{K'}) with uniaxial strains applied along the Ox-axis (transport direction).}
\label{fig_sim1}
\end{figure}
In addition to being attractive to electronics, graphene also shows remarkable mechanical properties. It has been experimentally demonstrated \cite{clee08} that graphene is amenable to a large strain of over 20$\%$. Actually, strain engineering offers a wide range of opportunities for modulating electronic properties of graphene nanostructures. In particular, the bandgap in armchair GNRs has a linear strain-dependence under weak strain, whereas it presents periodic oscillations at large strain \cite{ylu210}. For a zigzag GNR, strain can change the spin polarization at the ribbon edges, and therefore modulate its bandgap. In a 2D monolayer graphene, a finite gap can open under very large shear strain \cite{cocc10}. Otherwise, the bandgap may remain close to zero but the Dirac points are displaced \cite{pere09,huan10}. Many interesting electrical, optical, and magnetic properties induced by strain have been also reported, e.g., in refs. \cite{bunc07,pere09,kuma12,per010,pell10,guin10,tlow10,zhai11}.

In the present work, we focus on the effects of uniaxial strain applied in a 2D monolayer graphene sheet and explore the possibility to use a strained/unstrained hetero-channel (see Fig. 1) to achieve good operation of graphene FETs. Actually, local strain has been realized in experiments \cite{bunc07,bunc08,wbao09,tomo11,geor11,jlu012,kase13} and theoretically demonstrated to improve the electrical performance of various graphene devices \cite{pere09,ylu010,fuji10,juan11,baha13}. For instance, the local strain has been shown to enhance the ON current in a GNR tunneling FET by a factor of ten \cite{ylu010} and to fortify the transport gap in strained GNR junctions \cite{baha13}. We will demonstrate here that the use of strained heterostructure with a moderately small strain of $5 \%$ can greatly improve the performance of graphene FETs with a view to high ON/OFF ratio and good saturation of current in normal transistors; low subthreshold swing and strong non-linear effects in tunneling devices.

\section{Approach}

Our model is based on a nearest-neighbor tight-binding Hamiltonian \cite{hung09}. The application of a uniaxial strain along the Ox (transport) direction causes the following changes in the $C-C$ bond vectors \cite{baha13}:
\begin{eqnarray}
\left\{ \begin{array}{l}
 {r_x} \to \left( {1 + \sigma } \right){r_x} \\
 {r_y} \to \left( {1 - \nu \sigma } \right){r_y}
 \end{array} \right.
\end{eqnarray}
where $\sigma$ represents the strain and $\nu \simeq 0.165$ is the Poisson ratio \cite{blak70}. The hopping parameters between atoms \emph{i} and \emph{j} are defined as $t_{ij} = t_0 \exp\left[-3.37\left(r_{ij}/r_0 - 1\right)\right]$, where $r_{ij}$ is their bond length while the hopping energy $t_0 = -2.7$ $eV$ and the bond length $r_0 = 0.142$ $nm$ in the unstrained case. Therefore, there are two different hoping parameters $t_{1,2}$ along the armchair and zigzag directions, respectively, in the strained graphene (see in Fig. 1). This tight-binding Hamiltonian is solved in the ballistic limit using the Green's function technique, self-consistently with the Poisson's equation (for more details, see ref. \cite{hung12}). After the self-consistency is reached, the current is computed from the Landauer equation as
\begin{equation}
J = \frac{e}{{\pi h}}\int\limits_{BZ} {d{k_y}\int\limits_{ - \infty }^\infty  {d\epsilon\mathcal{T}\left( {\epsilon,{k_y}} \right)\left[ {{f_S}\left( \epsilon \right) - {f_D}\left( \epsilon \right)} \right]} }
\end{equation}
where $f_{S(D)} \left( \epsilon \right)$ is the source (drain) Fermi distribution function, $\mathcal{T}\left( {\epsilon,k_y}\right)$ is the transmission coefficient and the integral over the wave-vector $k_y$ is performed in the first Brillouin zone.

In our simulated devices schematized in Fig. 1, the gate of length 80 nm covers symmetrically both sides of the hetero-channel with a transition zone of $\sim$ 9 $nm$ between unstrained and strained sections. We assume the doping concentration $N_D$ = 10$^{13}$ cm$^{-2}$ in the source and drain access regions and the gate stack made of high-$\kappa$ insulator \cite{meri08}. We consider in this work two typical devices: normal FET (nFET) where the source and drain access regions are both $n$-doped and tunneling FET (TFET) where the drain is $p$-doped.

\section{Results and discussion}

\begin{figure}[!t]
\centering
\includegraphics[width=3.3in]{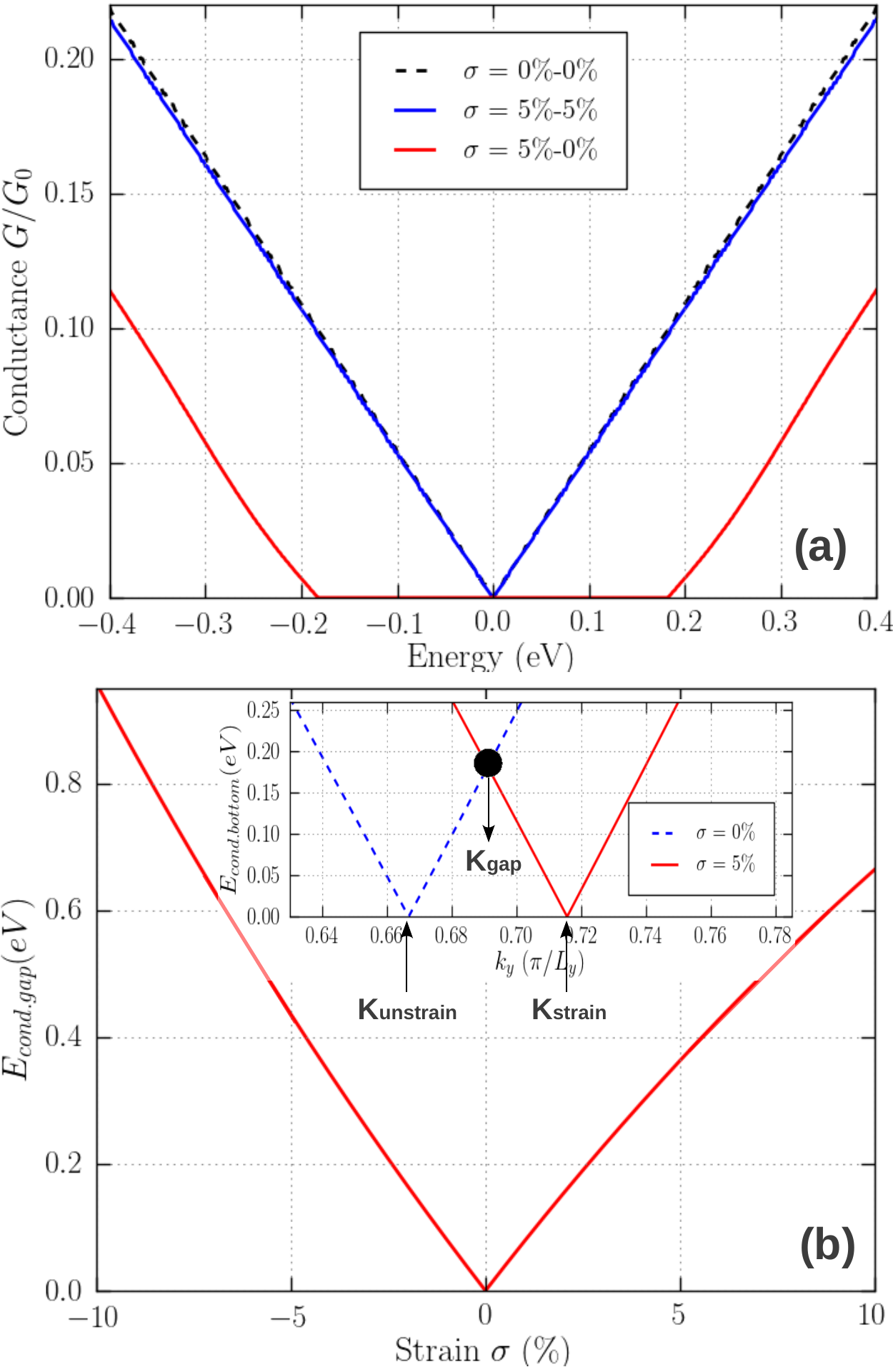}
\caption{Conductance ($G_0 \equiv e^2W/hL_y$) as a function of energy and evolution of the conduction gap in strained graphene heterostructures. The comparison of the conductance in unstrained, uniformly strained and strained heterostructures is shown in (a). The conduction gap as a function of strain in the heterostructures is displayed in (b) while the inset shows the bottom of conduction bands versus wave vector $k_y$.}
\label{fig_sim2}
\end{figure}
First, let us analyze the basic transport properties of strained/unstrained graphene heterostructures. Solving the tight binding Hamiltonian for strained graphene, we obtain the energy bands:
\begin{equation}
E^2 \left(\mathbf{k}\right) = t_1^2 + 4t_2^2 \cos^2 \left(\theta_y\right) \pm 4t_1t_2 \cos \left(\theta_x\right)\cos \left(\theta_y\right)
\end{equation}
where $\theta_{x(y)} = k_{x(y)}L_{x(y)}/2$ with the length $L_{x(y)}$ of unit cell as indicated in Fig. 1. Hence, the bandgap of uniformly strained graphene remains zero at $k_x = 2n\pi/L_x$ and $\cos \left(\theta_y\right) = \pm t_1/2t_2$ but the Dirac points are displaced under the effect of strain, as illustrated schematically in Fig. 1. This does not change the zero conduction-gap characteristic of graphene channel (see Fig. 2(a)) when we consider the conductance as a function of energy. However, in strained/unstrained heterostructures, we notice a very interesting phenomenon: the appearance of a finite conduction gap and this gap increases monotonically when increasing the strain (see Fig. 2(b)). This phenomenon is explained as follows. For a given $k_y$-mode, graphene has an energy gap defined as ${E_{gap}}\left( {{k_y}} \right) = 2\left| {2{t_2}\left| {\cos \left( {\frac{{{k_y}{L_y}}}{2}} \right)} \right| - {t_1}} \right|$. Because of the shift of Dirac points induced by strain as further illustrated by the plot of the conduction-band bottom versus $k_y$ in the inset of Fig. 2(b), there is always a finite energy gap $E_{gap}^{het} \left(k_y\right)$ of transmission in the strained heterostructure. The value of this gap is the maximum of $E_{gap} \left(k_y\right)$ in the strained and unstrained graphene sections. Therefore, a finite gap of conductance occurs (see Fig. 2(a)) and is determined by the minimum value of $E_{gap}^{het} \left(k_y\right)$ when varying $k_y$ in the whole Brillouin zone: ${E_{cond.gap}} = 2\left| \frac{{{t_1} - {t_2}}}{{{t_0} + {t_2}}} t_0 \right|$. This conduction gap is of course strain-dependent. For instance, it reaches about 360 $meV$ for $\sigma = 5$ $\%$ and can have a higher value for larger strains (see Fig. 2(b)). Similar properties of strained GNR junctions have also been discussed in ref. \cite{baha13}. This interesting feature is the key point suggesting that the performance of graphene FETs can be improved using a strained/unstrained hetero-channel.

\begin{figure}[!t]
\centering
\includegraphics[width=3.4in]{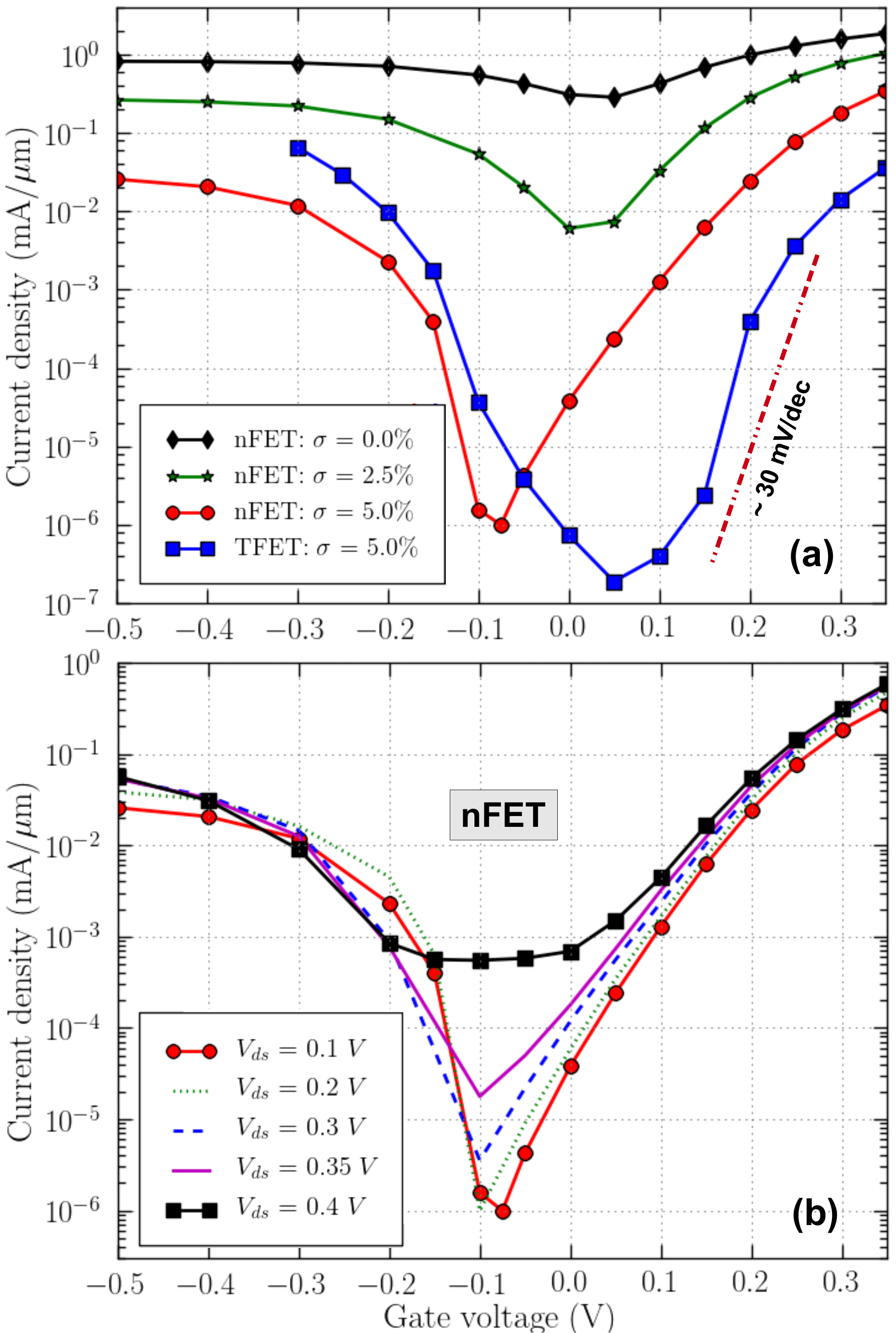}
\caption{$I-V_{gs}$ characteristics of simulated graphene FETs at $V_{ds} = 0.1$ $V$ with different strains (a) and for $\sigma = 5$ $\%$ with different $V_{ds}$ (b).}
\label{fig_sim3}
\end{figure}
\begin{figure*}[!t]
\centering
\includegraphics[width=6.8in]{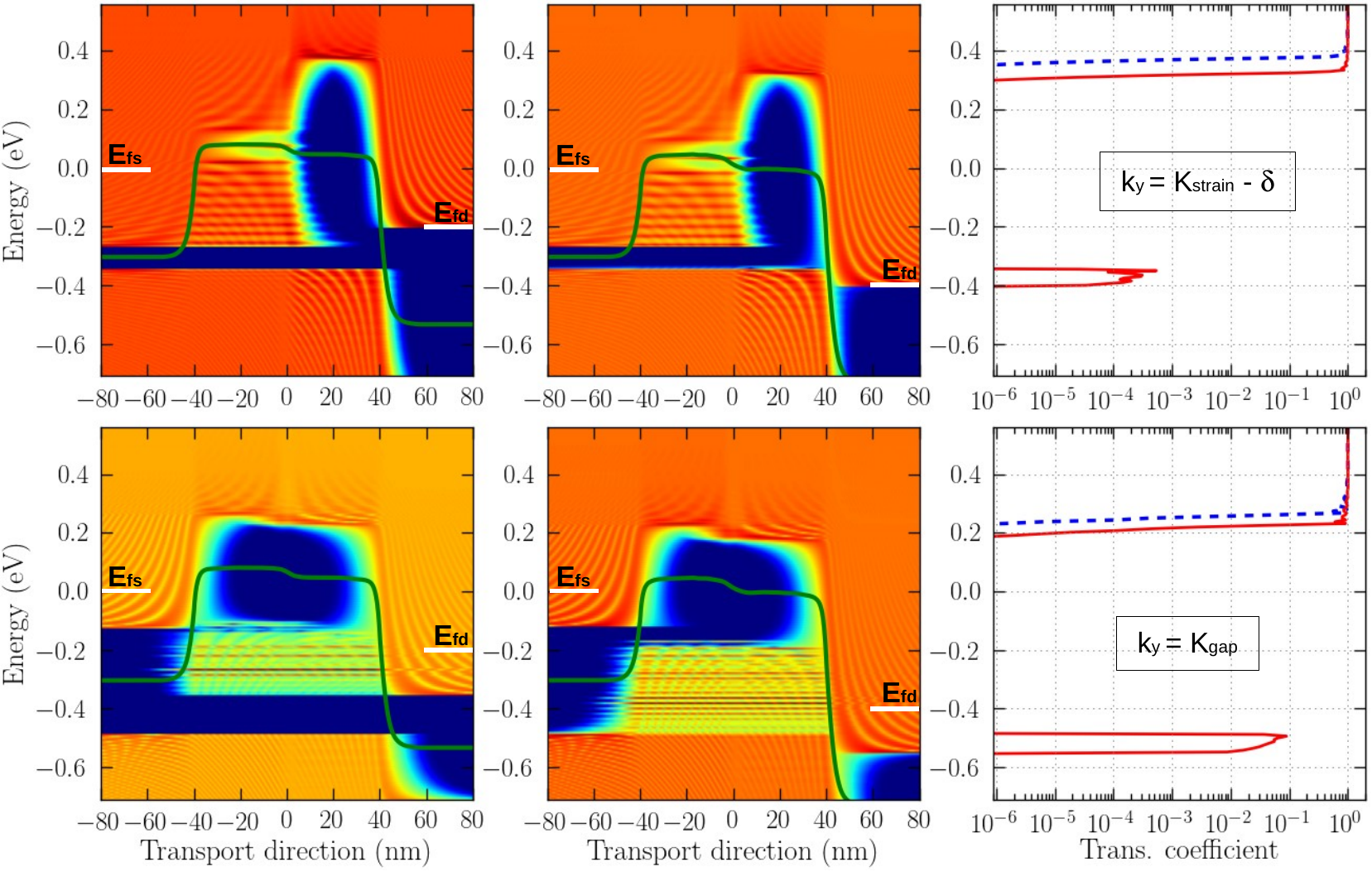}
\caption{Local density of states (left and middle panels) and corresponding transmission coefficient (right panels) for two different wave-vectors $k_y$ in the OFF state ($V_{gs} = -0.1$ $V$) of the nFETs studied in Fig. 3(b). The values $K_{strain}$ and $K_{gap}$ are defined as in Fig. 2(b), while $\delta = 0.005 \pi/L_y$. The drain voltage $V_{ds}$ = 0.2/0.4 $V$ in the left/middle panels, which corresponds to the dashed/solid lines in the right panels. The source and drain Fermi levels $E_{fs,fd}$ are indicated and the potential profile at the neutrality point is superimposed in the left and middle panels.}
\label{fig_sim4}
\end{figure*}
It has been shown \cite{alar13,berr13} that there are three components that contribute to the current in graphene nFETs: thermionic transmision in the energy regime higher than the gate-induced potential barrier, chiral tunneling through the barrier, and band-to-band tunneling taking place between the valence band in the source and the conduction band in the drain. While the band-to-band tunneling has an important contribution at high $V_{ds}$ which leads to a poor current saturation, the contribution of two other components results in a high OFF current (low ON/OFF current ratio) in gapless graphene devices. The use of a gapped graphene channel is hence necessary to solve these issues. We first focus on the possibility of using a strained/unstrained graphene channel to improve the performance of graphene nFETs. We find as displayed in Fig. 3(a) that when increasing the strain (or conduction gap), the OFF current is strongly reduced and, by defining arbitrarily the ON current as the current obtained at $V_{gs} = 0.35$ $V$, the ON/OFF current ratio significantly increases to over $10^5$ for $\sigma = 5$ $\%$. Such a high ratio seems to be at variance with what was shown for bilayer graphene \cite{fior09} and graphene nanomesh FETs \cite{berr13} where an energy gap of $\sim 300$ $meV$ is not enough to switch off the current efficiently, i.e., to obtain a large ON/OFF ratio. This may be because in these devices, the energy gap occurs locally in the gated region, and therefore is not sufficient to suppress the thermionic and/or band-to-band tunneling components. The situation here is different and can be explained as follows. In fact, the transport modes $k_y$ that contribute significantly to the current are the values $k_y$ in/around the range from $K_{unstrain}$ to $K_{strain}$ (see in the inset of Fig. 2(b)). Far from this range to the left and to the right, the energy gap ${E_{gap}}\left( {{k_y}} \right)$ in the strained (source and half of the gated zone) and unstrained (drain and half of the gated zone) graphene sections, respectively, is too large, which suppresses very strongly the corresponding current component even in the ON state. Close to $k_y = K_{unstrain}$ or $K_{strain}$, the energy gap mentioned is still large but the corresponding current component starts to have a significant contribution to the ON current while it remains almost negligible in the OFF state. For other modes $k_y$, e.g., around $K_{gap}$, the energy gap of $\sim E_{cond.gap}$, though smaller than the gap of modes discussed above, occurs along the whole channel and is thus enough to switch off the corresponding component of current in the OFF state as illustrated by the pictures of local density of states and transmission coefficient in Fig. 4 for $V_{ds} = 0.2$ $V$. The combined effect of these $k_y$-dependent energy gaps (discussed further below) makes the ON/OFF current ratio very large.

Next, we investigate the operation of graphene TFETs using the same strained hetero-channel. The conventional TFETs, besides being able to offer a high ON/OFF current ratio, have an additional advantage of having steep subthreshold swings lower than the thermally-limited value of 60 $mV/dec$ \cite{seab10}. This good performance, in principle, can not be achieved with a gapless channel as in unstrained graphene devices. Using the strained hetero-channel, we find as shown in Fig. 3(a) that an ON/OFF current ratio higher than $10^5$ is also achieved while the subthreshold swing can reach a value smaller than 30 $mV/de$c. All the results obtained demonstrate that strained hetero-channels are excellent candidates for graphene devices in digital applications.

We now go to explore the influence of $V_{ds}$. In Fig. 3(b), we display $I-V_{gs}$ curves obtained in graphene nFETs for different $V_{ds}$. Interestingly, it is shown that a very high ON/OFF current ratio can still be achieved for high $V_{ds}$ up to about 0.3 $V$. Beyond this value of $V_{ds}$, the OFF current increases more rapidly and the ON/OFF ratio is hence reduced, but it still reaches a value of $\sim 10^3$ at $V_{ds} = 0.4$ $V$. This can be understood by looking at the pictures of local density of states and transmission coefficient in Fig. 4. Again, for $k_y$ around $K_{gap}$, the energy gaps in the source, gated, and drain regions are merged together at $V_{ds}$ lower than $0.3$ $V$, which forms a large enough conduction gap to suppress strongly the corresponding current component. For $k_y$ around $K_{strain}$ (similarly, for $k_y$ around $K_{unstrain}$), the energy gaps in the gated and drain regions are also merged together, and thus efficiently suppress the current. This explains a small OFF current obtained for $V_{ds}$ lower than $0.3$ $V$. The situation suddenly changes at high $V_{ds}$ when the band-to-band tunneling comes into play. Indeed, as seen in Fig. 4 for $V_{ds} = 0.4$ $V$, the energy gap in the drain is separated from the others and the band-to-band tunneling has an important contribution to the current. This results in the rapid increase of the OFF current with $V_{ds}$ shown in Fig. 3(b).

\begin{figure}[!t]
\centering
\includegraphics[width=3.4in]{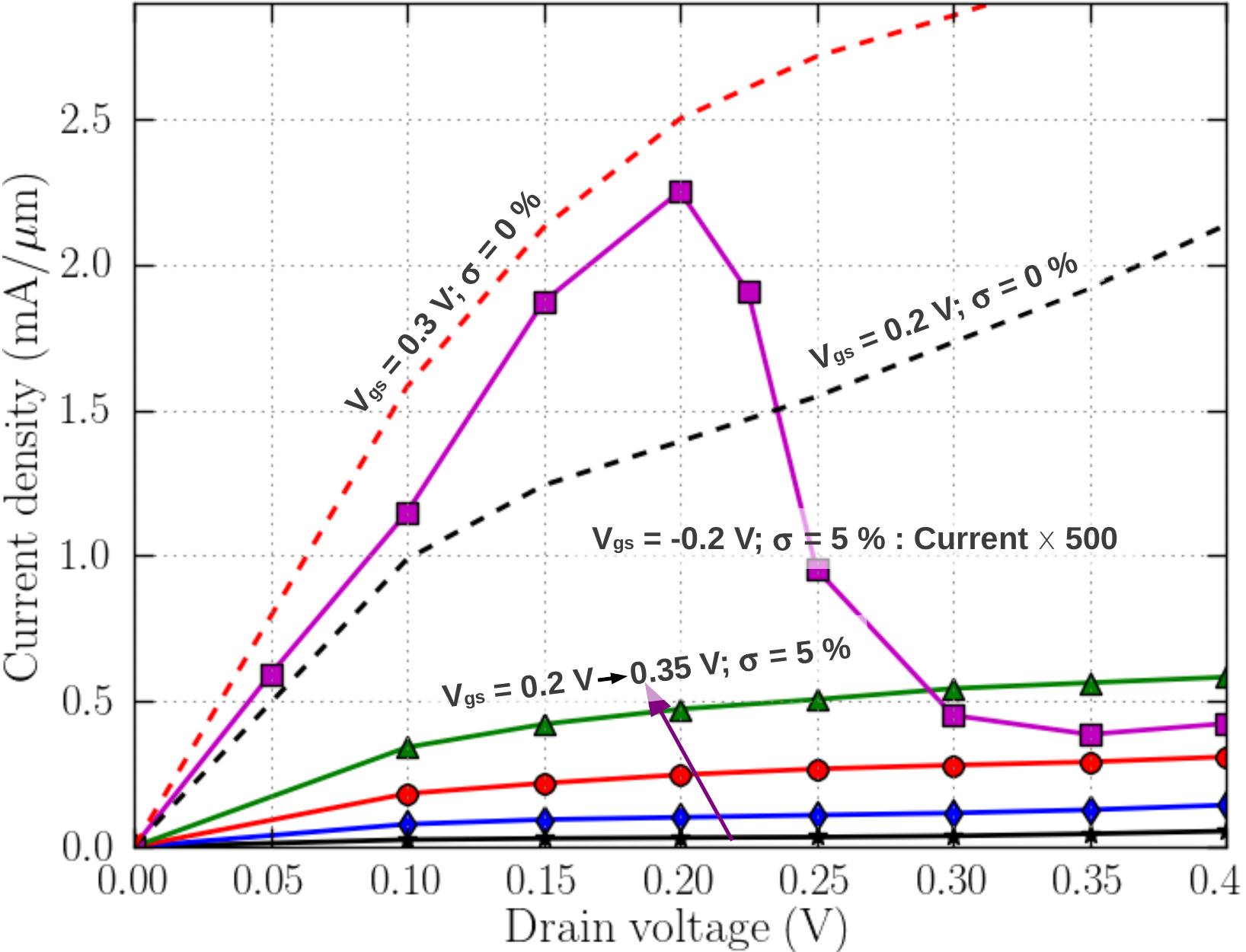}
\caption{$I-V_{ds}$ characteristics of simulated graphene nFETs with different $V_{gs}$: unstrained graphene channel ($\sigma = 0$ $\%$) and strained graphene heterostructure ($\sigma = 5$ $\%$). Note that the current at $V_{gs} = -0.2$ $V$ is multiplied by 500.}
\label{fig_sim5}
\end{figure}
In the negative range of $V_{gs}$, when the channel is $p$-doped, it has been shown in ref. \cite{alar13} that a pseudo-saturation of current or even a negative differential conductance (NDC) can be observed and was explained by the reduction of chiral tunneling when increasing $V_{ds}$ together with the limited contribution of thermionic/band-to-band currents. Similar effects are also observed in our simulated devices in the range of $V_{gs}$ from $-0.4$ $V$ to $-0.05$ $V$. Because of the strong effect of energy gaps mentioned, a strong NDC behavior with a peak-to-valley ratio of $\sim 6$ is even observed, as shown in the $I-V_{ds}$ curve presented in Fig. 5 for $V_{gs} = - 0.2$ $V$. The saturation behavior is also achieved in the $n$-branch (very positive $V_{gs}$), where the thermionic current is dominant (see the further discussions below). Additionally, it is remarkable to notice that in this regime, though reduced compared to the unstrained graphene FETs, the transconductance $g_m$ of the strained hetero-channel devices is still high. In particular, the transconductance has the value of $\sim$ 4.3 $mS/\mu m$ for $V_{gs} = V_{ds} = 0.3$ $V$, while it is about 13.9 $mS/\mu m$ in the unstrained case.
\begin{figure}[!t]
\centering
\includegraphics[width=3.4in]{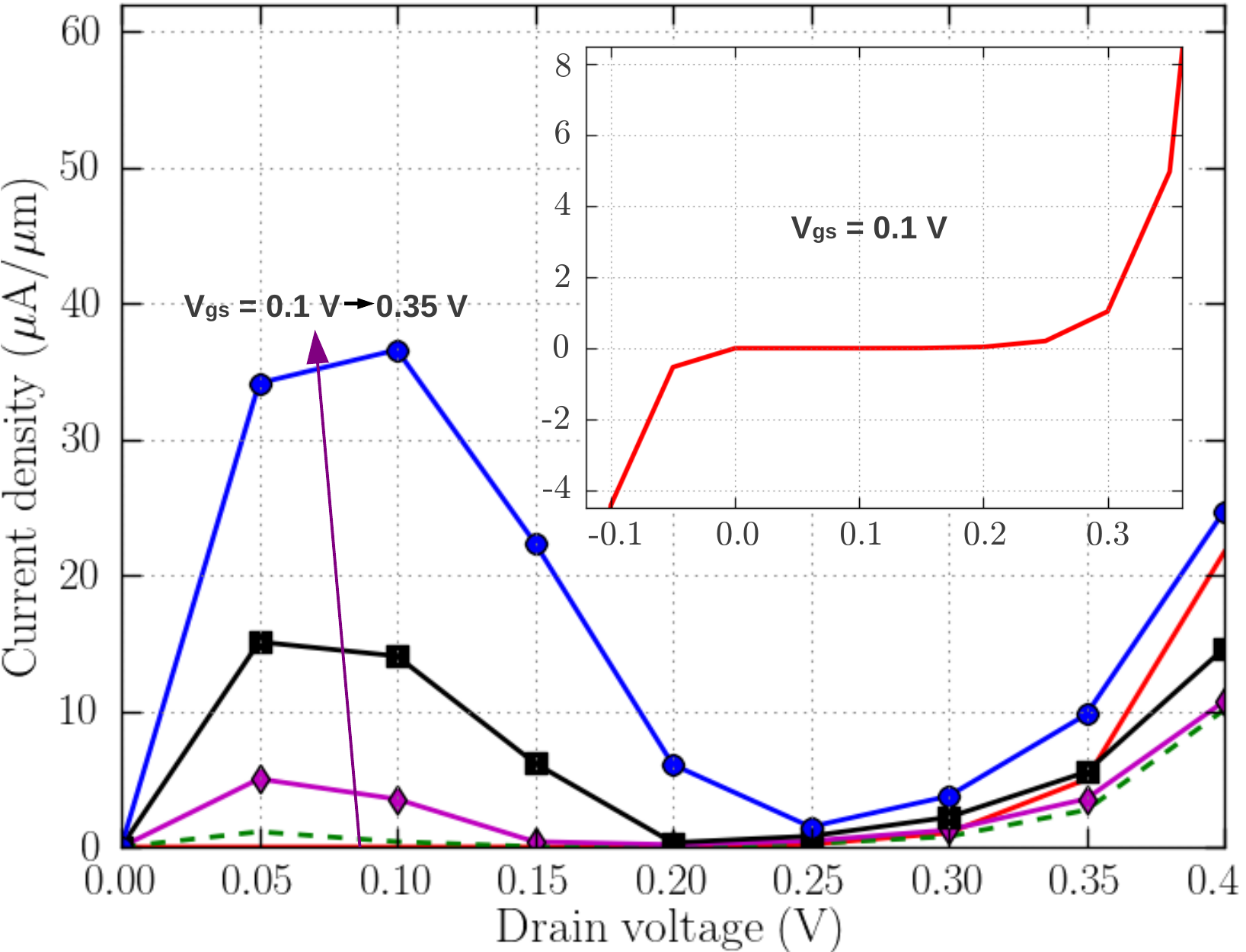}
\caption{$I-V_{ds}$ characteristics of simulated graphene TFETs ($\sigma = 5$ $\%$) for $V_{gs}$ ranging from 0.1 $V$ to 0.35 $V$. The inset shows the current with strong rectification at $V_{gs} = 0.1$ $V$.}
\label{fig_sim6}
\end{figure}

Combined with the high transconductance, the good saturation of current in the positive $V_{gs}$-range would be expected to result in a high value of voltage gain $A_v = g_m/g_d$, where $g_d$ is the output conductance. In Fig. 5, we display the current as a function of drain voltage in the $V_{gs}$-range from $0.2$ $V$ to 0.35 $V$. Indeed, the saturation behavior of current is significantly improved when using strained hetero-channels, compared to the unstrained graphene FETs. As a consequence of the reduced output conductance, the voltage gain at $V_{ds} = 0.3$ $V$ reaches the values of $\sim$ 9.5 and 18 for $V_{gs} = 0.2$ $V$ and 0.3 $V$, respectively, while it is only $\sim$ 2.8 and 4.7 in the unstrained device. To further evaluate the operation of the device in the high-frequency range, we estimate the cutoff frequency $f_T$ and the maximum oscillation frequency $f_{max}$ using the standard expressions $f_T = g_m/2\pi C_G$ and $f_{max} = f_T/\alpha$ with $\alpha = 2\sqrt{g_d(R_S+R_G)+g_mR_GC_{GD}/C_G}$ \cite{farm12}. $C_G$ and $C_{GD}$ are the total gate and gate-to-drain capacitances while $R_S$ and $R_G$ are the source and gate resistances, respectively. We take the values of $R_S$ and $R_G$ reported in refs. \cite{fxia11,farm12}, where $R_S$ is typically $\sim$ 100 $\Omega$ $\mu m$ and $R_G$ reasonably reaches a value of 200 $\Omega$ $\mu m$. For $V_{gs} = V_{ds} = 0.3$ $V$, we obtain $f_T \simeq 290$ GHz while $f_{max} \simeq 304$ GHz is even slightly larger than $f_T$. Similar behaviors have also been reported in some experiments \cite{farm12,ywu011}. The corresponding values obtained in the unstrained case are $f_T \simeq 898$ GHz and $f_{max} \simeq 356$ GHz. Thus, though reduced in the strained hetero-channel device, the transconductance and transition frequencies are still high. Together with the good current saturation and hence high voltage gain, our results demonstrate that the strained hetero-channel devices are also very promising for high-frequency applications.

Finally, we additionally found that the graphene TFETs can exhibit very strong non-linear effects on the $I-V_{ds}$ characteristics. In particular, two specific features, namely, gate controllable negative differential conductance and strong rectification of current, are observed, as shown in Fig. 6. Actually, the gate controllable NDC behavior in TFETs based on a gapped graphene channel has been discussed and explained in ref. \cite{hung12b} by assuming the possibility of controlling the gap profile in the gated zone. Here, our simulations show that the strong NDC behaviour can be achieved in the ON state with a peak-to-valley ratio of a few tens (e.g., 47 for $V_{gs} = 0.3$ $V$). Moreover, because of the strong effects of energy gaps, a finite gap of current occurs in/around the OFF state. This leads to high ON/OFF current ratio in a finite range of $V_{ds}$, similar to that in the graphene nFETs discussed above, with the possibility of tuning strongly the non-linearity of the $I-V_{ds}$ curves using the gate voltage. It also results in a strong effect of current rectification around $V_{ds} = 0$, as clearly seen in the inset of Fig. 6. These non-linear effects can offer great potential for applications such as oscillator, frequency multiplier, fast switching \cite{mizu95}, in non-Boolean logic circuits recently proposed \cite{gliu13}, and in rectifier circuits.

\section{Conclusion}

In summary, we have investigated the strain effects on the transport properties of unstrained/strained graphene heterostructures and considered the possibility of using such heterostructures to improve the performance of graphene transistors. It was shown that due to the deformation of electronic bandstructure, the mismatch between transport modes having different effective energy gaps in strained/unstrained sections results in a finite conduction gap in these graphene hetero-channels. We have demonstrated that this type of hetero-channels can improve very significantly the performance of graphene FETs for both digital and high-frequency applications. In particular, an ON/OFF current ratio higher than $10^5$ can be achieved. In graphene nFETs, while transconductance and transition frequencies are still high, the performance are additionally improved with respect to conventional graphene FETs in terms of current saturation and hence voltage gain. In graphene TFETs, low subthreshold swing and strong non-linear effects such as gate controllable NDC behavior and current rectification are observed. This type of strain-induced conduction-gap engineering could certainly be useful for improving performance of other devices based on graphene/graphene-like materials and for other applications.

\textbf{\textit{Acknowledgment.}} V. H. Nguyen and H. V. Nguyen thank Vietnam's National Foundation for Science and Technology Development (NAFOSTED) for financial support under Grant no. 103.02-2012.42. P. Dollfus acknowledges the French ANR for financial support under the projects NANOSIM-GRAPHENE (Grant no. ANR-09-NANO-016) and MIGRAQUEL (Grant no. ANR-10-BLAN-0304).

\end{document}